\title{The Universe as Quantum Computer}
\author{
        Seth Lloyd\\
        Department of Mechanical Engineering\\
        Massachusetts Institute of Technology\\
        Cambridge MA 02139 USA
}

\documentclass[12pt]{article}

\begin{document}
\maketitle

\begin{abstract}

This article reviews the history of digital computation, and investigates
just how far the concept of computation can be taken.  
In particular, I address the
question of whether the universe itself is in fact a giant computer,
and if so, just what kind of computer it is.  I will show that
the universe can be regarded as a giant quantum computer.  The
quantum computational model of the universe explains a variety
of observed phenomena not encompassed by the ordinary laws
of physics.  In particular, the model shows that the
the quantum computational universe automatically 
gives rise to a mix of randomness and order,
and to both simple and complex systems. 

\end{abstract}

\section{Introduction}

It is no secret that over the last fifty years the world has
undergone a paradigm shift in both science and technology.
Until the mid-twentieth century, the dominant paradigm
in both science and technology was that of energy: over the
previous centuries, the laws of physics had been developed
to understand the nature of energy and how it could be
transformed.   In concert with progress in physics,
the technology of the industrial revolution put the new
understanding of energy to use for manufacturing and
transportation.  In the mid-twentieth century, 
a new revolution began.  This revolution was based not
on energy, but on information.  The new science of
information processing, of which Turing was one of
the primary inventors, spawned a technology
of information processing and computation.  This technology
gave rise to novel forms and applications of
computation and communication.   The rapid spread
of information processing technologies, in turn, has ignited
an explosion of scientific and social inquiry. 
The result is a paradigm shift of how we think about
the world at its most fundamental level. \index{paradigm shift}  Energy
is still an important ingredient of our understanding
of the universe, of course, but information has attained
a conceptual and practical status equal to -- and frequently
surpassing -- that of energy.  Our new understanding of
the universe is not in terms of the driving power of force
and mass.  Rather, the world we see around us arises from
a dance between equal partners, information and energy, 
where first one takes the lead and then the other.  The bit meets the erg,
and the result is the universe.  

At bottom, the information that makes up the universe is not
just ordinary classical information (bits).  Rather, it 
is quantum information (qubits).   Consequently, the computational
model that applies the universe at its smallest and most fundamental
level is not conventional digital computation, but quantum
computation \cite{Lloyd1}. \index{quantum computation}  The strange and weird
aspects of quantum mechanics infect the universe at its
very beginning, and -- as will be seen -- provide the mechanism
by which the universe generates its peculiar mix 
of randomness, order, and complexity \index{complexity}.

\section{Digital computation before Turing}

Before describing how the universe can be modeled as a quantum
computer, and how that quantum computational model of the
universe explains previously unexplained features, we
review computation and computational models of the universe
in general.

Alan Turing \index{Turing, Alan} 
played a key role in the paradigm shift from energy
to information: his development of a formal theory of digital computation
made him one of the most influential mathematicians
of the twentieth century.   It is fitting, therefore, to praise him. 
Curiously, however, Turing's seminal role in a global
scientific and technological revolution also leads to
the temptation to over-emphasize his contributions.  We human
beings have a sloppy, if not outright bad habit of assigning
advances to a few `great men.'  I call this habit the Pythagoras
syndrome,  \index{Pythagoras syndrome} 
after the tendency in the western world to assign
all pre-fourth century B.C.E. mathematics to Pythagoras \index{Pythagoras} 
without regard to actual origins.
In evaluating Turing's contributions, we should be careful
not to fall victim to the Pythagoras syndrome, if only to give
full credit to his actual contributions, which were specific and great.

Computing machines are not a modern invention \cite{Cunningham}: the abacus was
invented in Babylon more than four thousand years ago. \index{abacus}  Analog,
geared, information processing mechanisms were developed in China
and Greece thousands of years ago, and attained considerable
sophistication in the hands of medieval Islamic philosophers. 
John Napier's \index{Napier, John}
seventeenth century mechanical implementation
of logarithms (`Napier's bones') was the precursor of the slide
rule.  The primary inventor of the modern digital computer, however,
was Charles Babbage. \index{Babbage, Charles}   
In 1812, Babbage had the insight that the calculations
carried out by mathematicians could be broken down into sequences
of less complicated steps, each of which could be carried out
by a machine \cite{Babbage} 
-- note the strong similarity to Turing's insight
into the origins of the Turing machine more than a century later.
The British government fully appreciated the potential impact
of possessing a mechanical digital computer, and funded Babbage's
work at a high level.  During the 1820s he designed and attempted to build
a series of prototype digital computers that he called 
`difference engines.' \index{difference engine}  
Nineteenth century manufacturing tolerances
turned out to be insufficiently precise to construct the 
the all-mechanical difference engines, however.  
The first large-scale computing project consumed over seventeen
thousand pounds sterling of the British taxpayers' money, 
a princely expenditure for pure research at the time.
Like many computing projects since, it failed.

Had they been constructed, difference engines would have
been able to compute general polynomial functions, but they
would not have been capable of what Turing termed universal
digital computation.  After the termination of funding for
the difference engine project, Babbage turned his efforts to
the design of an `analytic engine.' \index{analytic engine}
Programmed by punched
cards like a Jacquard loom, the analytic engine would have been
a universal digital computer.  The mathematician Ada Lovelace
\index{Lovelace, Ada}
devised a program for the analytic engine to compute Bernoulli
numbers, thereby earning the title of the world's first
computer programmer.

The insights of Babbage and Lovelace occurred more than
a century before the start of the information processing
revolution.   Turing was born in the centenary of the
year in which Babbage had his original insight.  
The collection in which this paper appears
could equally be dedicated to the two-hundredth anniversary
of Babbage's vision.  But scientific history 
is written to celebrate winners (see Pythagoras, above).  
Turing `won' the title of the inventor
of the digital computer because his insights played
a direct role in the vision of the creators of the
first actual physical computers in the mid-twentieth
century.  The science fiction genre known as `steampunk' speculates 
how the world might have evolved if nineteenth century technology
had been up to the task of constructing the difference
and analytical engines. \index{steampunk}  
(Perhaps the best-known example
of the steampunk genre is William Gibson \index{Gibson,
William} and Bruce Sterling's \index{Sterling, Bruce}
novel, `The Difference Engine' \cite{Gibson}.)

The mathematical development
of digital logic did not occur until after Babbage's 
mechanical development.
It was not until the 1830s and 1840s that the British
logician Augustus de Morgan \index{de Morgan, Augustus}
and the mathematician
George Boole \index{Boole, George} developed the bit-based logic on which
current digital computation is based. \index{Boolean logic}  
Indeed, had Babbage
been aware of this development at the time, the physical
construction of the difference and analytic engines might
have been easier to accomplish, as Boolean, bit-based
operations are more straightforward to implement mechanically
than base-ten operations.  The
relative technological simplicity of bit-based operations
would play a key role in the development of electronic
computers.


By the time that Turing began working on the theory of computation,
Babbage's efforts to construct actual digital computers were
a distant memory.  Turing's work had its direct intellectual
antecedents in the contentious arguments on the logical and
mathematical basis of set theory that were stirred up at the
beginning of the twentieth century.  At the end of the nineteenth
century, the German mathematician
David Hilbert \index{Hilbert, David}
proposed an ambitious programme to axiomatize
the whole of mathematics.  In 1900, he famously formulated this programme at
the International Congress of Mathematicians in Paris as a challenge
to all mathematicians -- a collection of twenty three problems whose
solution he felt would lead to a complete, axiomatic theory not
just of mathematics, but of physical reality.  Despite or because
of its grand ambition to establish the logical foundations of 
mathematical thought, cracks began to appear in Hilbert's programme 
almost immediately.   The difficulties arose at the most fundamental
level, that of logic itself.  Logicians
and set theorists such as Gottfried Frege \index{Frege, Gottfried}
and Bertrand Russell \index{Russell, Bertrand}
worked for decades to make set theory consistent, but the net result
of their work was call into question the logical foundations of 
set theory itself.  In 1931, just when the efforts of mathematicians
such as John von Neumann \index{von Neumann, John}
had appeared to patch up those
cracks,  Kurt G\"odel \index{G\"odel, Kurt}
published his beautiful but disturbing
incompleteness theorems, \index{incompleteness theorem} 
showing that any system of logic that
is powerful enough to describe the natural numbers is fundamentally
incomplete in the sense that there exist well-formulated proposition
within the system that cannot be resolved using the system's axioms
\cite{Goedel}.
By effectively destroying Hilbert's
programme, G\"odel's startling result jolted the mathematical
community into novel ways of approaching the very notion of
what logic was.     

\section{Digital computation concurrent with Turing}

Turing's great contribution to logic can be thought of as
the rejection of logic as a Platonic ideal, and the
redefinition logic as a {\it process}.  
Turing's famous paper of 1936, `On Computable Numbers 
with an application to the  {\it Entscheidungsproblem},'
showed that the process of performing Boolean logic could
be implemented by an abstract machine \cite{Turing}, subsequently called
a Turing machine. \index{Turing machine}  Turing's machine was an abstraction
of a mathematician performing a calculation by thinking and
writing on pieces of paper.  The machine has a `head' to do the thinking,
and a `tape' divided up in squares to form the machine's memory.  The head
has a finite number of possible states, as does each square.  At each
step, in analogy to the mathematician looking at the piece of
paper in front of her, the head reads the state of the square on which it sits.
Then, in analogy to thinking and writing on the paper,
the head changes its state and the state of the square.    
The updating occurs as a function of the head's 
current state and the state of the square. 
Finally, in analogy to the mathematician either taking up a new
sheet of paper or referring back to one on which she has previously
written, the head moves one square to the left of right,
and the process begins again.

Turing was able to show that such machines were very powerful
computing devices in principle.  In particular, he proved
the existence of `universal' Turing machines, which were
capable of simulating the action of any other Turing machine,
no matter how complex the actions of its head and squares.
Unbenownst to Turing, the American mathematician Alonzo
Church \index{Church, Alonzo}
had previously arrived at a purely formal logical
description of the idea of computability \cite{Church}, the so-called
Lambda calculus. \index{Lambda calculus}  
At the same time as Turing, Emil Post \index{Post, Emil}
devised a mechanistic treatment of logical problems.
The three methods were all formally equivalent, but
it was Turing's that proved the most accessible.

Perhaps the most fascinating aspect of Turing's mechanistic
formulation of logic was how it dealt with the self-contradictory and
incomplete aspects of logic raised by G\"odel's incompleteness
theorems.  G\"odel's theorems arise from the ability of logical
systems to have self-referential statements -- they are a formalization
of the ancient `Cretan liar paradox,' in which a statement
declares itself to be false.  If the statement is true, then 
it is false; if it is false, then it is true.  Regarding
proof as a logical process, G\"odel restated the paradox
as a statement that declares that it can't be proved to be
true.  There are two possibilities.  If the statement is false,
then it can be proved to be true -- but if a false statement
can be proved to be true, then the entire system of logic
is inconsistent.  If the statement is true, then it can't
be proved to be true, and the logical system is incomplete.

In Turing's formulation, logical statements about proofs
are translated into actions of machines.  The self-referential
statements of G\"odel's incompleteness theorems then translate
into statements about a universal Turing machine that is programmed
to answer questions about its own behavior.  In particular,
Turing showed that no Turing machine could answer the question
of when a Turing machine `halts' -- i.e., gives the answer
to some question. \index{halting problem}   If such a machine existed, then it
would be straightforward to construct a related machine
that halts only when it fails to halt.  In other words,
the simplest possible question one can ask of a digital
computer -- whether it gives any output at all -- cannot
be computed!

The existence of universal Turing machines, together with
their intrinsic limitation due to self-contradictory
behavior as in the halting problem, has profound consequences
for the behavior of existing computers.  In particular,
current electronic computers are effectively universal
Turing machines.  Their universal nature expresses itself
in the fact that it is possible to write software that
can be compiled to run on any digital computer, no matter
whether it is made by HP, Lenovo, or Apple.  The power
of universal Turing machines manifests itself in the remarkable
power and flexibility of digital computation.  This
power is expressed in the so-called Church-Turing hypothesis,
which states that any effectively calculable function can
be computed using a universal Turing machine.  
The intrinsically self-contradictory nature of Turing machines
and the halting problem manifest themselves in the intrinsically
annoying and frustrating behavior of digital computers --
the halting problem implies that there is no systematic
way of debugging a digital computer.  No matter what one
does, there will always be situations where the computer 
exhibits surprising and unexpected behavior (e.g.,
the `blue screen of death').

Concurrent with the logical, abstract development of the
notion of computation, including Turing's abstract machine,
engineers and scientists were pursuing the construction
of actual digital computers.  In Germany in 1936, 
Konrad Zuse \index{Zuse, Konrad} designed the Z-1,
a mechanical calculator whose program was written 
on perforated 35mm film.  In 1937, Zuse expanded the
design to allow universal digital computation {\it a la} Turing.
When completed in 1938, the Z-1 functioned poorly due
to mechanical imprecision, the same issue that plagued
Babbage's difference engine more than a century earlier.
By 1941, Zuse had constructed the Z-3, an electronic
computer capable of universal digital computation. 
Because of its essentially applied nature, and because
it was kept secret during the second world war, Zuse's
work received less credit for its seminal nature than
was its due (see the remark above on winner's history).

Meanwhile, in 1937, Claude Shannon's \index{Shannon, Claude}
MIT master's thesis,
 `A Symbolic Analysis of Relay and Switching Circuits,'
showed how any desired Boolean function -- including
those on which universal digital computation could be based --
could be implemented using electronic switching circuits \cite{Shannon}.
This work had a profound influence on the construction
of electronic computers in the United States and Great
Britain over the next decades.

\section{Digital computation post-Turing}

Turing's ideas on computation had immediate impact on the
construction of actual digital computers.  While doing
his Ph.D. at Princeton in 1937, Turing himself constructed
simple electronic models of Turing machines.  The real
impetus for the development of actual digital computers
came with the onset of the second world war.  Calculations
for gunnery and bombing could be speeded up electronically.
Most relevant to Turing's work, however, was the use
of electronic calculators for the purpose of cryptanalysis 
During the war, Turing became the premier code-breaker for
the British cryptography effort.  The first large-scale
electronic computer, the Colossus, was constructed to
aid this effort.   In the United States, IBM constructed
the Mark I at Harvard, the second programmable computer
after Zuse's Z-3, and used it to perform ballistic
calculations.  Zuse himself had not remained idle:
he created the world's first computer start-up,
designed the follow-up to the Z-3, the Z-4, and wrote
the first programming language.  The end of the war saw  
the construction of the Electronic Numerical Integrator
and Computer, or ENIAC. 

To build a computer requires and architecture.
Two of the most influential proposals for computer architectures
at the end of the war were the Electronic Discrete Variable 
Automatic Computer, or EDVAC, authored by von Neumann,
and Turing's Automatic Computing Engine, or ACE.  
Both of these proposals implemented what is called
a `von Neumann' computer architecture, in which program
and data are stored in the same memory bank.  Stored program
architectures were anticipated by Babbage, implicit in Turing's
original paper, and had been developed previously by
J. Presper Eckert \index{Eckert, J. Presper}
and John Mauchly \index{Mauchly, John} in their design
for the ENIAC.   The Pythagoras syndrome, however,
assigns their development to von Neumann, who himself
would have been unlikely to claim authorship.

This ends our historical summary of conventional digital
computation.  The last half century 
has seen vast expansion of devices, techniques, and architectures
notably the development of the transistor and integrated circuits.
But the primary elements of computation -- programmable systems to perform
digital logic -- were all in place by 1950.  

\section{The computing universe}

The physical universe bears little resemblance to the collection
of wires, transistors, and electrical circuitry that make up a
conventional digital computer.  How then, can one claim
that the universe is a computer?  The answer lies in the
definition of computation, of which Turing was the primary
developer.  According to Turing, a universal digital computer
is a system that can be programmed to perform any desired 
sequence of logical operations.  Turing's invention of
the universal Turing machine makes this notion precise.
The question of whether the universe is itself a universal
digital computer can be broken down into two parts:
(I) Does the universe compute? and (II) Does the universe
do nothing more than compute?  More precisely,
(I) Is the universe capable of performing universal
digital computation in the sense of Turing?  That is, can
the universe or some part of it be programmed to
simulate a universal Turing machine?  (II) Can a universal
Turing machine efficiently simulate the dynamics of
the universe itself?  

At first the answers to these questions might appear,
straightforwardly, to be Yes.  When we construct electronic
digital computers, we are effectively programming some
piece of the universe to behave like a universal digital
computer, capable of simulating a universal Turing machine.
Similarly, the Church-Turing hypothesis implies, that
{\it any} effectively calculable physical dynamics -- including
the known laws of physics, and any laws that may be discovered
in the -- can be computed using a digital computer.  

But the straightforward answers are not correct.  First,
to simulate a universal Turing machine requires 
a potentially infinite supply of memory space.
In Turing's original formulation, when a Turing machine
reaches the end of its tape, new blank squares can always
be added: the tape is `indefinitely extendable.'
Whether the universe that we inhabit
provides us with indefinitely extendable memory is 
an open question of quantum cosmology, and will be discussed
further below.  So a more accurate answer to the first
question is `Maybe.'   The question of whether or not infinite
memory space is available is not so serious, as one can formulate
notions of universal computation with limited memory.  After
all, we treat our existing electronic computers as universal
machines even though they have finite memory (until, of course,
we run out of disc space!).  The fact that we possess computers
is strong empirical evidence that laws of physics support universal
digital computation.

The straightforward answer to question (II) is more doubtful.
Although the outcomes of any calculable laws
of physics can almost certainly be simulated on a universal Turing machine,
it is an open question whether this simulation can be
performed {\it efficiently} in the sense that a relatively
small amount of computational resources are devoted to
simulating what happens in a small volume of space and time.
The current theory of computational complexity suggests
that the answer to the second question is `Probably not.'   

An even more ambitious programme for the computational
theory of the universe is the question of architecture.
The observed universe possesses the feature that the laws
of physics are local -- they involve only interactions
between neighboring regions of space and time.  Moreover,
these laws are homogeneous and isotropic, in that they
appear to take the same form in all observed regions of
space and time.  The computational version of a homogeneous system
with local laws is a cellular automaton, a digital system
consisting of cells in regular array. \index{cellular automaton}  
Each cell possesses
a finite number of possible states, and is updated as
a function of its own state and those of its neighbors.
Cellular automata were proposed by von Neumann and by
the mathematician Stanislaw Ulam \index{Ulam, Stanislaw}
in the 1940s, and used
by them to investigate mechanisms of self-reproduction 
\cite{von Neumann 1}.
Von Neumann and Ulam showed that cellular automata were
capable of universal computation in the sense of Turing.
In the 1960s, Zuse and computer scientist Edward Fredkin 
\index{Fredkin, Edward}
proposed that cellular automata could be used as the
basis for the laws of physics -- i.e., the universe is
nothing more or less than a giant cellular automaton \cite{Zuse}.
More recently, this idea was promulgated by Stephen
Wolfram. \index{Wolfram, Stephen}

The idea that the universe is a giant cellular automaton
is the strong version of the statement that the universe
is a computer.  That is, not only does the universe compute,
and only compute, but also if one looks at the `guts'
of the universe -- the structure of matter at its smallest
scale -- then those guts consist of nothing more than
bits undergoing local, digital operations.  The strong
version of the statement that the universe is a computer
can be phrased as the question, (III) `Is 
the universe a cellular automaton?'  As will now
be seen, the answer to this question is No.
In particular, basic facts about quantum mechanics prevent
the local dynamics of the universe from being reproduced
by a finite, local, classical, digital dynamics.

\section{Classical digital devices can't reproduce quantum mechanics 
efficiently}

Quantum mechanics is the physical theory that describes 
how systems behave at their most fundamental scales. \index{quantum 
mechanics} 
It was studying von Neumann's book \cite{von Neumann 2}
{\it The mathematical foundations
of quantum mechanics} that inspired Turing to work on
mathematics \cite{Hodges}.    (In particular, Turing was interested in
reconciling questions of determinism and free will
with the apparently indeterministic nature of
quantum mechanics.)  Quantum mechanics is well-known for
exhibiting strange, counter-intuitive features.  Chief
amongst these features is the phenomenon known as entanglement,
which Einstein \index{Einstein, Albert} 
termed `spooky action at a distance' ({\it spukhafte
Fernwirkung}).  In fact, entanglement does not engender non-locality
in the sense of non-local interactions or superluminal communication.  
However, a variety of theorems from von Neumann to Bell \index{Bell,
John} and
beyond show that the types of correlations implicit in entanglement
cannot be described by classical local models involving
hidden variables \cite{Zurek}.  In particular, such quantum correlations
cannot be reproduced by local classical digital models
such as cellular automata.   
Non-local classical hidden variable models can reproduce
the correlations of quantum mechanics, but only at the
by introducing either superluminal communication,
or a very large amount of classical information
to reproduce the behavior of a single quantum bit.
Accordingly, the answer
to question (III), is the universe a cellular automaton,
is `No.'

The inability of classical digital systems to cope with
entanglement also seems to prevent ordinary computers from
simulating quantum systems efficiently.  Merely to represent
the state of a quantum system with $N$ subsystems, e.g.,
$N$ nuclear spins, requires $O(2^N)$ bits on a classical
computer.  To represent how that state evolves requires
the exponentiation of a $2^N$ by $2^N$ matrix.  Although
it is conceivable that exponential compression techniques
could be found that would allow a classical computer
to simulate a generic quantum system efficiently,
none are known.  So the currently accepted answer
to question (II), can a Turing machine simulate
a quantum system efficiently, is `Probably not.'

\section{Quantum computing}

The difficulty that classical computers have reproducing
quantum effects makes it difficult to sustain
the idea that the universe might at bottom be a classical
computer.   Quantum computers, by definition, are
good at reproducing quantum effects, however \cite{Chuang}.  Let's
investigate the question of whether the universe might be,
at bottom, a quantum computer \cite{Lloyd1}.  

A quantum computer is a computer that uses quantum effects
such as superposition and entanglement to perform computations
in ways that classical computers cannot. \index{quantum computer}  
Quantum computers
were proposed by Paul Benioff \index{Benioff, Paul}
in 1980 \cite{Benioff}.  The notion of a quantum
Turing machine that used quantum superposition to perform
computations in a novel way was proposed by David Deutsch \index{Deutsch,
David} in
1985 \cite{Deutsch}.  
For a decade or so, quantum computation remained something
of a curiosity.  No one had a particularly good application
for them, and no one had the least idea how to build them.
The situation changed in 1994, when Peter Shor \index{Shor, Peter} 
showed that a relatively modestly sized quantum computer, containing
a few thousand logical quantum bits or `qubits,'
and capable of performing around a million coherent operations,
could be used to factor large numbers and so break
public key cryptosystems such as RSA \cite{Shor}.  The previous year,
Lloyd \index{Lloyd, Seth} 
had showed how quantum computers could be constructed
by applying electromagnetic pulses to arrays of coupled quantum
systems \cite{Lloyd2}.  The resulting parallel quantum computer is in
effect a quantum cellular automaton.  In 1995, Ignacio Cirac
\index{Cirac, Ignacio} and Peter Zoller \index{Zoller, Peter}
showed how ion traps could be used to
implement quantum computation \cite{Cirac}.  

Since then, a wide variety of designs for quantum computers
have been proposed.  Further quantum algorithms have been developed, 
and prototype quantum computers have been constructed and used to
demonstrate simple quantum algorithms.  This allows us to begin
addressing the question of whether the universe is a quantum
computer.  If we `quantize'
our three questions, the first one, (Q1) `Does the universe allow
quantum computation?' has the provisional answer, `Yes.'
As before, the question of whether the universe affords
a potentially unlimited supply of quantum bits remains open.
Moreover, it is not clear that human beings currently possess
the technical ability to build large scale quantum computers
capable of code breaking.  However, from the perspective
of determining whether the universe supports quantum computation,
it is enough that the laws of physics allow it.

Now quantize the second question.  (Q2) `Can a quantum
computer efficiently simulate the dynamics of the universe?'
Because they operate
using the same principles that apply to nature at fundamental
scales, quantum computers -- though difficult to construct --
represent a way of processing information that is closer
to the way that nature processes information at the microscale.
In 1982, Richard Feynman \index{Feynman, Richard}
suggested that quantum devices could
function as quantum analog computers to simulate the dynamics
of extended quantum systems \cite{Feynman}.  In 1996, Lloyd developed a quantum
algorithm for implementing such universal quantum simulators 
\cite{Lloyd3}. 
The Feynman-Lloyd results show that, unlike classical computers,
quantum computers can simulate efficiently any quantum system
that evolves by local interactions, including for example
the standard model of elementary particles.  While no
universally accepted theory of quantum gravity currently 
exists, as long as that theory involves local interactions
between quantized variables, then it can be efficiently
simulated on a quantum computer.  So the answer to
the quantized question 2 is `Yes.'

There are of course subtleties to how a quantum computer
can simulate the known laws of physics.  Fermions supply
special problems of simulation, which however can be overcome.
A short-distance (or high-energy) cutoff in the dynamics
is required to insure that the amount of quantum
information required to simulate local dynamics is finite.
However, such cutoffs -- for example, at the Planck scale --
are widely expected to be a fundamental feature of nature.

Finally, we can quantize question three: (Q3) `Is the universe
a quantum cellular automaton?'  While we cannot unequivocally 
answer this question in the affirmative, we note that
the proofs that show that a quantum computer can simulate
any local quantum system efficiently immediately imply that
any homogeneous, local quantum dynamics, such as that given
by the standard model and (presumably) by quantum gravity,
can be directly reproduced by a quantum cellular automaton.
Indeed, lattice gauge theories, in Hamiltonian form, map
directly onto quantum cellular automata.  Accordingly,
all current physical observations are consistent with
the theory that the universe is indeed a quantum
cellular automaton.

\section{The universe as quantum computer}

We saw above that basic aspects of quantum mechanics, such as
entanglement, make it difficult to construct a classical computational
model of the universe as a universal Turing machine or
a classical cellular automaton.  By contrast, the power
of quantum computers to encompass quantum dynamics allows
the construction of quantum computational models of the
universe.  In particular, the Feynman-Lloyd construction
allows one to map any local, homogeneous quantum dynamics
directly onto a quantum cellular automaton.  

The immediate question is `So what?'  Does the fact
that the universe is observationally indistinguishable
from a giant quantum computer tell us anything new or
interesting about its behavior?  The answer to this
question is a resounding `Yes!'  
In particular, the
quantum computational model of the universe answers
a question that has plagued human beings ever since
they first began to wonder about the origins of the
universe, namely, Why is the universe so ordered and
yet so complex \cite{Lloyd1}?  

The ordinary laws of physics tell
us nothing about why the universe is so complex.
Indeed, the complexity of the universe is quite
mysterious in ordinary physics.  The reason is
that the laws of physics are apparently quite simple. 
The known ones can be written down on the back of a tee shirt.
Moreover, the initial state of the universe appears also 
to have been simple. Just before the big bang, the universe was
highly flat, homogeneous, isotropic, and almost entirely
lacking in detail.  Simple laws and simple initial conditions
should lead to states that are, in principle, themselves
very simple.  But that is not what we see when we look 
out the window.  Instead we see vast variety and detail --  
animals and plants, houses and humans, and overhead, at night,
stars and planets wheeling by.  Highly complex systems and
behaviors abound.  The quantum computational model of the universe
not only explains this complexity: it requires it to exist.

To understand why the quantum computational model necessarily
gives rise to complexity, consider the old story of monkeys
typing on typewriters. \index{typing monkeys}  
The original version of this story
was proposed by the French probabilist \'Emile Borel,
\index{Borel, Emile}
at the beginning of the twentieth century (for a detailed account
of the history of typing monkeys see \cite{Lloyd1}).  Borel imagined
a million typing monkeys ({\it singes dactylographes}) and pointed
out that over the course of single year, the monkeys had
a finite chance of producing all the texts in all the libraries
in the world.   He then immediately noted that with very high
probability, they would would produce nothing but gibberish.

Consider, by contrast, the same monkeys typing into computers.
Rather than regarding the monkeys random scripts as mere texts,
the computers interpret them as programs, sets of instructions
to perform logical operations.  At first it might seem that the
computers would also produce mere gibberish -- `garbage in, garbage
out,' as the programmer's maxim goes.  While it is true that
many of the programs might result in garbage or error messages,
it can be shown mathematically that the monkeys have a relatively
high chance of producing complex, ordered structures.  The reason
is that many complex, ordered structures can be produced from
short computer programs, albeit after lengthy calculations. 
Some short program will instruct the computer to calculate
the digits of $\pi$, for example, while another will cause
it to produce intricate fractals.  Another will instruct the
computer to evaluate the consequences of the standard model
of elementary particles, interacting with gravity, 
starting from the big bang.  A particularly brief program instructs
the computer to prove all possible theorems.
Moreover, the shortest programs to produce these complex structures
are necessarily {\it random}.  If they were not, then there would
be an even shorter program that could produce the same structure.
So the monkeys, by generating random programs, are producing 
exactly the right conditions to generate structures of arbitrarily 
great complexity.  

For this argument to apply to the universe itself, two ingredients
are necessary -- first, a computer, and second, monkeys. 
But as shown above, the universe itself is indistinguishable
from a quantum computer.  In addition, quantum fluctuations -- 
e.g., primordial fluctuations in energy density --
automatically provide
the random bits that are necessary to seed the quantum computer
with a random program.  That is, quantum fluctuations are the
monkeys that program the quantum computer that is the universe.
Such a quantum computing universe {\it necessarily} generates
complex, ordered structures with high probability.

\section{Conclusions}

This article reviewed the history of computation with the
goal of answering the question, `Is the universe a computer?'
The inability of classical digital computers to reproduce
quantum effects efficiently makes it implausible that the
universe is a classical digital system such as a cellular
automaton.  However, all observed phenomena are consistent
with the model in which the universe is a quantum computer,
e.g., a quantum cellular automaton.  The quantum computational
model of the universe explains previously unexplained features,
most importantly, the co-existence in the universe of randomness 
and order, and of simplicity and complexity.

\end{document}